\documentstyle[12pt]{article}
\begin {document}

\title {THE RELATIVISTIC TRANSFORMATION TO ROTATING FRAMES}

\author{L. Herrera\thanks{ Also at UCV, Caracas, Venezuela; e-mail address:
lherrera@gugu.usal.es}\\Area de F\'\i
sica Te\'orica. Facultad de Ciencias.\\ Universidad de Salamanca. 37008
Salamanca, Espa\~na.}

\date{}
\maketitle

\begin{abstract}
We present a critical review of the relativistic rotation
transformation of Trocheris and Takeno.
A new transformation is proposed which is free from the
drawbacks of the former. Some applications are presented.
\end{abstract}

\newpage

\section{Introduction}

The problem of determining the relativistic transformation
to rotating frames has been the subject of lengthy discussions
since the appearance of the special theory of relativity in 1905.

After a first wave of publications on this issue \cite{Eh}--\cite{La}, the
well known set of equations
\begin{equation}
\rho' = \rho \quad ; \quad \phi' = \phi-\omega t \quad ; \quad
z' = z \quad ; \quad t' = t
\label{pri}
\end{equation}
was adopted, relating the non-rotating coordinate system of the frame $S$
(in cylindrical coordinates) to the coordinates $t', \rho', z', \phi'$ of
the frame $S'$ rotating uniformly about $z-$axis in the $(\rho, \phi)$ plane.

More recently the tetrad formalism \cite{Sy},\cite{Pi} has been applied to
the study
of rotating observers \cite{LiNi}--\cite{Co} where uniform rotation is
defined according
to eq.(\ref{pri}).

It is not difficult to realize the inconvenients implied by eq.(\ref{pri}),
from the relativistic point of view. In fact, the absolute character of time
$(t=t')$ (of purely Galilean origin) is difficult to reconcile with special
relativity (clocks in $S'$ move with respect to clocks in $S$), even if we
are aware of the fact that $t$ is the coordinate time
and not proper time.

Furthermore, eq.(\ref{pri}) implies that the Galilean
composition law of velocities applies \cite{Tr}--\cite{KiKr}.

In order to overcome difficulties steaming from the Galilean character of
(\ref{pri}), a different transformation law was independently proposed by
Trocheris \cite{Tr} and Takeno \cite{Ta}.

In cylindrical coordinates the Trocheris-Takeno (TT) transformation reads
$$
\rho' = \rho \qquad ; \qquad z' = z
\label{roztt}
$$
\begin{equation}
\phi' = \left[\phi \, \cosh{\lambda} - t \, \frac{c}{\rho}\,
\sinh{\lambda}\right]
\label{tt}
\end{equation}
$$
t' = \left[t \, \cosh{\lambda} - \phi \, \frac{\rho}{c} \, \sinh{\lambda}\right]
$$
with
$$
\lambda \equiv \frac{\rho \omega}{c}
$$
where primes correspond to the rotating frame.

It can be easily seen \cite{Tr,Ta} that (\ref{tt}) takes the ordinary
form of Lorentz transformation, leads to the relativistic law of
composition of velocities and yields for the velocity of a fixed
point in $S'$, the expression

\begin{equation}
v=c \,\tanh{\lambda}
\label{v}
\end{equation}
sending to infinity the ``light cylinder''.
Of course, up to first order in $\lambda$, we recover
$$
\phi'=\phi-\omega t
$$

Although all these features are quite desirable in a self-consistent
relativistic transformation, those two references (\cite{Tr},\cite{Ta})
have been largely overlooked, except for a series of papers by
Kichenassamy and Krikorian \cite{KiKr}.

The purpose of this work is twofold, on one hand we want to call
the attention to TT transformation and on the other, to propose some
modifications of it, in order to avoid some undesirable consequences
following from its application.

The motivation to undertake such endeavor is provided not only by
the evident academic interest of the problem but also by the fact
that navigation and time transfer systems are now being contemplated
with sub-nanosecond time accuracy (see \cite{Ba} and references therein).
For such systems, terms of order in $\lambda$ higher than one will be required.

In the next section we shall rederive the TT transformation following
the Takeno approach, and will discuss about its properties. Other
transformations obtained within the same scheme will also be considered.
In section 3, a new transformation will be proposed and some applications
will be presented. Finally some conclusions are included in the last section.

\section{The TT transformation}
\subsection{Its derivation}
We shall now discuss with some detail the derivation of TT
transformation as given by Takeno. Since this reference is
almost half century away in the past, we shall try to make
this section self-consistent, providing as much details as
possible and following closely, with minor changes, the
original notation.

Thus, we start by assuming that $U(\omega)$, which defines
a uniform rotation constitutes a group with continuous parameter
$\omega$. Let us next consider an inertial system $S$ with respect
to which a system $S'$ makes a $U(\omega)$, and a system $S''$ making a
$U(\bar\omega)$ with respect to $S'$.

Then consider a fixed point $A'$ in $S'$.
Denoting its three dimensional velocity $(v^x,v^y,0)$ (as measured by $S$)
and its speed by $v$, we have
\begin{equation}
v^x = - v \sin{\phi} = - y \xi \quad ; \quad v^y = v \cos{\phi} = x \xi
\label{vs}
\end{equation}
where $\xi = v/\rho$ is the ``angular velocity'' of $A'$ as measured by $S$,
and $x,y \, (\rho,\phi)$ its cartesian (cylindrical) coordinates.

It is further assumed that $\xi$ is a function of $\rho$ and $\omega$, and
for sufficiently small $\lambda \, (\lambda = \rho \omega/c)$ it coincides
with $\omega$ up to first order in $\lambda$.

We shall further consider two other points $A$ and $A''$ at rest in $S$
and $S''$ respectively and which coincide with $A'$ at some instant of time.

Then denoting by $\overline{v}, \overline{\xi}$ and
$\overline{\overline{v}}, \overline{\overline{\xi}}$ the
velocity and the
``angular velocity'' of $A''$ with respect to $S'$ and $A''$ relative to $S$,
respectively, we shall demand that $\overline{\overline{v}}$ is obtained
from $v$ and
$\overline{v}$ according to the relativistic composition law, i.e.
\begin{equation}
\overline{\overline{v}}=\frac{v+\overline{v}}{1+ v \overline{v}/c^2}
\label{com}
\end{equation}
which implies
\begin{equation}
\overline{\overline{\xi}}=\frac{\xi+\overline{\xi}}{1+ \rho^2 \xi
\overline{\xi}/c^2}
\label{comx}
\end{equation}

Next, constraining the rotation to the $(\rho,z)$ plane is natural to assume
\begin{equation}
\rho' = \rho \quad ; \quad z' = z
\label{roz}
\end{equation}

We shall now consider infinitesimal transformations assuming $\omega$ to be
small. Then, up to linear terms in $\omega$, we may write
\begin{equation}
x'^{\alpha} = x^\alpha + \omega \eta^\alpha \qquad \qquad
(x^{0,1,2,3}=t,\rho,\phi,z)
\label{prior}
\end{equation}
then from (\ref{roz}) it follows that
\begin{equation}
\eta^1 = \eta^3 = 0
\label{eta}
\end{equation}

Using
\begin{equation}
\overline{\overline\xi} = \frac{d\phi}{dt} \qquad;\qquad \overline\xi =
\frac{d\phi'}{dt'}
\label{bx}
\end{equation}
one finds without difficulty
\begin{equation}
\overline{\xi}=\left\{\overline{\overline{\xi}} + \omega
\left(\overline{\overline{\xi}}
\, \frac{\partial \eta^2}{\partial\phi} +
\frac{\partial \eta^2}{\partial t}\right)\right\}/
\left\{1 + \omega \left(\overline{\overline{\xi}}
\, \frac{\partial \eta^0}{\partial\phi} +
\frac{\partial \eta^0}{\partial t}\right)\right\}
\label{bbx}
\end{equation}
Next, since $\omega$ is small, we may write, up to terms linear in $\omega$
\begin{equation}
\xi=\omega
\label{om}
\end{equation}
then we obtain from (\ref{comx})
\begin{equation}
\overline{\xi}=\frac{\overline{\overline{\xi}}-\omega}{1-\rho^2
\overline{\overline{\xi}}\omega/c^2}
\label{baxi}
\end{equation}
and comparing (\ref{baxi}) with (\ref{bbx}), the following equations follow
 \begin{equation}
\frac{\partial \eta^2}{\partial \phi}=0 \quad;\quad
\frac{\partial \eta^2}{\partial t}=-1 \quad;\quad
\frac{\partial \eta^0}{\partial \phi}=-\frac{\rho^2}{c^2} \quad;\quad
\frac{\partial \eta^0}{\partial t}=0
\label{par}
\end{equation}
which may be integrated to obtain
\begin{equation}
\eta^2 = -t \qquad;\qquad \eta^0 = - \frac{\rho^2}{c^2} \, \phi
\label{int}
\end{equation}
where the initial conditions $t'=\phi'=0$ for $t=\phi=0$ have been used.

Let us now get back to equation (\ref{prior}), taking derivative
with respect to $\omega$ and using (\ref{int}), we have
\begin{equation}
\frac{d\phi'}{d\omega} = -t \qquad;\qquad
\frac{dt'}{d\omega} = - \frac{\rho^2}{c^2} \, \phi
\label{dfit}
\end{equation}

Also, because of (\ref{prior}) and (\ref{int}), we may write
\begin{equation}
\phi'=\phi-\omega t \qquad;\qquad t'=t-\frac{\rho^2 \omega}{c^2} \, \phi
\label{fit'}
\end{equation}
or, solving for $\phi$ and $t$
\begin{equation}
\phi=\frac{\phi'+\omega t'}{1-\omega^2 \rho^2/c^2} \qquad;\qquad
t=\frac{t'+ (\omega \rho^2/c^2)\, \phi'}{1-\omega^2 \rho^2/c^2}
\label{fit}
\end{equation}

However, since (\ref{prior}) is defined up to linear terms in $\omega$,
it is meaningful to neglect quadratic terms in (\ref{fit}) and write
\begin{equation}
\phi=\phi' + \omega t' \qquad;\qquad t=t' + \frac{\omega \rho^2}{c^2} \phi'
\label{neg}
\end{equation}
thus, eq.(\ref{dfit}) becomes
\begin{equation}
\frac{d\phi'}{d\omega} = - \left(t' + \frac{\omega \rho^2}{c^2} \phi'\right)
\qquad;\qquad
\frac{dt'}{d\omega} = - \frac{\rho^2}{c^2} \left(\phi'+\omega t'\right)
\label{dneg}
\end{equation}

Before proceeding further with the integration of (\ref{dneg})
the following observations are of order:
\begin{enumerate}
\item Takeno neglects the linear term in $\omega$ in (\ref{dneg}),
therefore his equations are:
\begin{equation}
\frac{d\phi'}{d\omega} = - t'
\qquad;\qquad
\frac{dt'}{d\omega} = - \frac{\rho^2}{c^2} \phi'
\label{Ta}
\end{equation}
which may be integrated at once, using the condition $\phi'=\phi ; t'=t$ for
$\omega=0$, to obtain (\ref{tt}).
\item It is important to observe that the inverse transformation
is obtained from (\ref{tt}) by changing $\omega \rightarrow -\omega$.
\item For a point at rest in $S'$, we have from (\ref{tt})
\begin{equation}
\phi=\frac{ct}{\rho} \tanh{\lambda} + \frac{a}{\cosh{\lambda}}
\label{fi}
\end{equation}
where we have put $\phi'=constant=a$, then
\begin{equation}
\xi=\frac{c}{\rho} \tanh{\lambda}
\label{xi}
\end{equation}
and therefore
\begin{equation}
v=c\tanh{\lambda}
\label{ve}
\end{equation}
thus a point co-rotating with $S'$ will move with speed approaching
$c$, only in the limit $\rho \rightarrow \infty$.
\end{enumerate}

Let us now integrate (\ref{dneg}) with the term linear in $\omega$
included. A simple calculation with the condition $\phi'=\phi; t'=t$ for
$\omega=0$, yields
\begin{equation}
\phi'=e^{-\lambda^2/2} \left[\phi \cosh{\lambda} - t \, \frac{c}{\rho}
\sinh{\lambda}\right]
\label{fip}
\end{equation}
\begin{equation}
t'=e^{-\lambda^2/2} \left[t \cosh{\lambda} - \phi \, \frac{\rho}{c}
\sinh{\lambda}\right]
\label{tp}
\end{equation}
or, solving for $\phi$ and $t$
\begin{equation}
\phi=e^{\lambda^2/2} \left[\phi' \cosh{\lambda} + t' \, \frac{c}{\rho}
\sinh{\lambda}\right]
\label{fi}
\end{equation}
\begin{equation}
t=e^{\lambda^2/2} \left[t' \cosh{\lambda} + \phi' \, \frac{\rho}{c}
\sinh{\lambda}\right]
\label{t}
\end{equation}

Observe that now, unlike the TT case , eqs.(\ref{fi}),(\ref{t}) are not
obtained from (\ref{fip}),(\ref{tp}) by changing $\omega \rightarrow -\omega$
and $\phi'^{\longrightarrow}_{\longleftarrow} \phi$ ;
$t'^{\longrightarrow}_{\longleftarrow} t$.

Let us now retrace all steps leading from (\ref{prior}) to (\ref{dneg}),
but for the inverse transformation (i.e. changing $\omega \rightarrow -\omega$
and $\phi'^{\longrightarrow}_{\longleftarrow} \phi$ ;
$t'^{\longrightarrow}_{\longleftarrow} t$). It is easily seen that the
following
equations result
\begin{equation}
\frac{d\phi}{d\omega} =  \left(t - \frac{\omega \rho^2}{c^2} \phi\right)
\qquad;\qquad
\frac{dt}{d\omega} =  \frac{\rho^2}{c^2} \left(\phi - \omega t\right)
\label{res}
\end{equation}
As expected, of course, (\ref{res}) may be obtained simply by changing
$\omega \rightarrow -\omega$
and $\phi'^{\longrightarrow}_{\longleftarrow} \phi$ ;
$t'^{\longrightarrow}_{\longleftarrow} t$, in (\ref{dneg}).

If we now solve (\ref{res}) without the terms in $\omega$ on the right of
equations, we obtain at once
\begin{equation}
\phi= \left[\phi' \cosh{\lambda} + t' \, \frac{c}{\rho} \sinh{\lambda}\right]
\label{fib}
\end{equation}
\begin{equation}
t= \left[t' \cosh{\lambda} + \phi' \, \frac{\rho}{c} \sinh{\lambda}\right]
\label{tb}
\end{equation}
which is just the TT transformation for the unprimed quantities, and which
follows from (\ref{tt}) by solving for $\phi$ and $t$ or simply by changing
$\omega \rightarrow -\omega$
and $\phi'^{\longrightarrow}_{\longleftarrow} \phi$ ;
$t'^{\longrightarrow}_{\longleftarrow} t$.

However if we solve (\ref{res}) with all their terms, the result is
\begin{equation}
\phi= e^{-\lambda^2/2} \left[\phi' \cosh{\lambda} + t' \, \frac{c}{\rho}
\sinh{\lambda}\right]
\label{fia}
\end{equation}
\begin{equation}
t= e^{-\lambda^2/2} \left[t' \cosh{\lambda} + \phi' \, \frac{\rho}{c}
\sinh{\lambda}\right]
\label{ta}
\end{equation}
or solving for $\phi'$ and $t'$
\begin{equation}
\phi'= e^{\lambda^2/2} \left[\phi \cosh{\lambda} - t \, \frac{c}{\rho}
\sinh{\lambda}\right]
\label{fiap}
\end{equation}
\begin{equation}
t'= e^{\lambda^2/2} \left[t \cosh{\lambda} - \phi \, \frac{\rho}{c}
\sinh{\lambda}\right]
\label{tap}
\end{equation}

Two facts deserve to be stressed:
\begin{enumerate}
\item As it was the case for (\ref{fip})--(\ref{t}), the expressions
for unprimed (primed) coordinates do not follow from the expressions
for primed (unprimed) coordinates, by changing
$\omega \rightarrow -\omega$
and $\phi'^{\longrightarrow}_{\longleftarrow} \phi$ ;
$t'^{\longrightarrow}_{\longleftarrow} t$.
This is a desirable feature of both transformations ((\ref{fip})--(\ref{t})
and (\ref{fia})--(\ref{tap})) since, unlike the case of Lorentz transformations
between two inertial frames, now $S$ and $S'$ are physically different (one is
inertial, whereas the other is not).
\item Transformations (\ref{fip})--(\ref{t}) are obviously different  from
(\ref{fia})--(\ref{tap}). This is to be expected from the point 1 above,
and the fact that (\ref{fip}),(\ref{tp}) ((\ref{fi}),(\ref{t}))
may be obtained from (\ref{fia}),(\ref{ta}) ((\ref{fiap}),(\ref{tap}))
(and viceversa) by changing
$\omega \rightarrow -\omega$
and $\phi'^{\longrightarrow}_{\longleftarrow} \phi$ ;
$t'^{\longrightarrow}_{\longleftarrow} t$. Also observe that
 Takeno approach
is, strictly speaking, consistent only up to terms linear in $\omega$.
It is obvious that TT, (\ref{fip})--(\ref{t}) and (\ref{fia})--(\ref{tap})
coincide up to that order and furthermore, primed and unprimed quantities
are, up to that order, obtained from each other by
$\omega \rightarrow -\omega$
and $\phi'^{\longrightarrow}_{\longleftarrow} \phi$ ;
$t'^{\longrightarrow}_{\longleftarrow} t$.
\end{enumerate}
To summarize: The TT transformation follows unambiguosly from the
Takeno procedure if linear terms in $\omega$, in eq.(\ref{dneg})
(or (\ref{res})) are neglected. Otherwise, the method does not provide
a unique set of transformations. We shall come back to this point latter,
now let us study some properties of TT transformation.

\subsection{Some properties of TT transformation}
In order to put in evidence some problems related to TT transformation,
let us calculate some metric components in the primed (rotating) system.
Thus, from
\begin{equation}
g_{t't'} = \frac{\partial x^{\alpha}}{\partial t'} \,
\frac{\partial x^{\beta}}{\partial t'} \,
g_{\alpha \beta}
\label{gtp}
\end{equation}
we easily obtain
\begin{equation}
g_{t't'} = c^2
\label{g}
\end{equation}

On the other hand, we know from general relativity that the relation between
the proper time interval $d\tau$ and the coordinate time interval $dt'$ is
given by
\begin{equation}
d\tau = \frac{1}{c} \sqrt{g_{t't'}} dt'
\label{tautp}
\end{equation}
which means that
\begin{equation}
d\tau = dt'
\label{tat}
\end{equation}
However, different observers in $S'$, located along a $\phi'=constant$,
$z'=constant$, line, moves with different velocities with respect to
each other (increasing with $\rho$ according to (\ref{v})) and
therefore their respective clocks should run differently,
specifically, as we move in the direction of increasing $\rho$,
they should go slower \cite{ABS}.

This is by the way the behaviour obtained from the ``Galilean''
transformation (\ref{pri}). Indeed we obtain in this case
\begin{equation}
g_{t't'}=c^2 - \omega^2 \rho^2
\label{gG}
\end{equation}
which means because of (\ref{tautp})
\begin{equation}
d\tau = \sqrt{1 - \frac{\omega^2 \rho^2}{c^2}} dt'
\label{tatG}
\end{equation}

Next, we obtain for $g_{\phi'\phi'}$ and $g_{t'\phi'}$
\begin{equation}
g_{\phi'\phi'} = - \rho^2 \qquad;\qquad g_{t'\phi'} = 0
\label{gfit}
\end{equation}
Therefore the length of a circle of radius $\rho$, measured in $S'$
becomes \cite{LaLi}
\begin{equation}
l'_{\phi} =  \sqrt{- g_{\phi'\phi'}} \, \int_0^{2\pi}{d\phi}
\label{li}
\end{equation}
or
\begin{equation}
l'_{\phi} = 2 \pi \rho
\label{l}
\end{equation}
which is obviously the same result obtained in $S$. Since observers
in both frames are in relative motion, one should expect, intuitively,
$l'_\phi > 2\pi\rho$ \cite{Ri} (see also the
letter of Einstein to J.Petzoldt, quoted in \cite{St}).. This is in fact
what one obtains
from the ``Galilean'' transformation (\ref{pri}), which yields \cite{LaLi}
\begin{equation}
l'_\phi = \frac{2\pi\rho}{\sqrt{1-\frac{\omega^2 \rho^2}{c^2}}} > 2\pi \rho
\label{lG}
\end{equation}

Let us now turn to transformations (\ref{fip})--(\ref{t}) and
(\ref{fia})--(\ref{tap}), to see what results do they provide
for proper time intervals and length of circles in the $(z,\rho)$
plane.

\subsection{Some properties of transformation (25)--(28)}
First of all observe that in this case, as in the TT case, the velocity of
a point at rest in $S'$, as measured in $S$,
is given by the expression (\ref{ve}).

Next, it follows from (\ref{fi}),(\ref{t}), that
\begin{equation}
g_{t't'} = c^2 e^{\lambda^2}
\label{gpp}
\end{equation}
which for $\lambda<<1$ gives, up to second order in $\lambda$
\begin{equation}
g_{t't'} \approx c^2 \left(1+\frac{\omega^2 \rho^2}{c^2}\right)
\label{gso}
\end{equation}
in evident contradiction with (\ref{gG}).

On the other hand, from
\begin{equation}
g_{\phi'\phi'} = - \rho^2 e^{\lambda^2}\qquad;\qquad g_{t'\phi'} = 0
\label{gg}
\end{equation}
it follows that the length of a circle of radius $\rho$, as measured in $S'$,
in given by
\begin{equation}
l'_{\phi'} = 2 \pi \rho e^{\lambda^2/2}
\label{lphi}
\end{equation}
which for $\lambda<<1$ gives, up to terms of order $\lambda^2$
\begin{equation}
l'_{\phi'} = 2 \pi \rho \left(1+\frac{\omega^2 \rho^2}{2c^2}\right)
\label{ll2}
\end{equation}
which coincides with the ``Galilean'' expression (\ref{lG}) expanded
up to the indicated order.

Thus, the transformation (\ref{fip})--(\ref{t}) provides a solution
to the problem of the length of the circle but gives an unsatisfactory
answer to the question of proper time interval.

Let us now consider the transformation (\ref{fia})--(\ref{tap}).

\subsection{Some properties of transformation (32)--(35)}
As in the previous case, the expression for the velocity of a
point at rest in $S'$ is given by (\ref{ve}).

Next, it follows from (\ref{fia}), (\ref{ta}) that
\begin{equation}
g_{t't'}=c^2e^{-\lambda^2}
\label{tptp}
\end{equation}
which yields, up to order $\lambda^2$
\begin{equation}
g_{t't'} \approx c^2 \left(1-\frac{\omega^2\rho^2}{c^2}\right)
\label{orl}
\end{equation}
in agreement with the result obtained from the ``Galilean''
transformation (\ref{pri}) (see eq.(\ref{gG})).

However, from
\begin{equation}
g_{\phi'\phi'}=-\rho^2 e^{-\lambda^2}\qquad ; \qquad
g_{t'\phi'}=0
\label{fifi}
\end{equation}
one obtains
\begin{equation}
l'_{\phi}=2 \pi \rho e^{-\lambda^2/2}
\label{lon}
\end{equation}
or, up to terms of order $\lambda^2$
\begin{equation}
l'_{\phi}\approx 2 \pi \rho \left(1-\frac{\omega^2\rho^2}{2c^2}\right)
< 2\pi \rho
\label{lc}
\end{equation}
in contradiction with (\ref{lG}).

Therefore, in this case the proper time problem is satisfactorily solved,
whereas the calculation of the circle's length leads to a wrong result. So,
it happens that each set of transformations solves correctly a problem which
is not solved by the other set. Based on this observation and on the fact,
mentioned above, that the Takeno procedure leads unambiguously to a set of
transformations (TT), only when the term linear in $\omega $, appearing in
(\ref{neg}) (or (\ref{res})) is neglected, we shall propose now a modified
TT transformation which is free from the objections brought up above.

\section{The modified TT transformation}
The results of previous section suggest the following set of
transformations (MTT)
\begin{equation}
t = \left(e^{-\lambda^2/2} \cosh{\lambda}\right) t' +
\left(e^{\lambda^2/2} \frac{\rho'}{c} \sinh{\lambda}\right) \phi'
\label{MTTt}
\end{equation}
\begin{equation}
\phi = \left(e^{-\lambda^2/2} \frac{c}{\rho'} \sinh{\lambda}\right)  t'+
\left(e^{\lambda^2/2} \cosh{\lambda}\right) \phi'
\label{MTTf}
\end{equation}
or, solving for $t'$ and $\phi'$
\begin{equation}
t' = e^{\lambda^2/2} \left[\left(\cosh{\lambda}\right)  t -
 \left(\frac{\rho}{c} \sinh{\lambda}\right) \phi \right]
\label{MTTtp}
\end{equation}
\begin{equation}
\phi' = e^{-\lambda^2/2} \left[ \left(\cosh{\lambda}\right) \phi -
\left(\frac{c}{\rho} \sinh{\lambda} \right) t\right]
\label{MTTfp}
\end{equation}
MTT coincides with TT except for the exponential factors and,
as it is seen from (\ref{MTTfp}), gives for the velocity of a
point at rest in $S'$, the same expression (\ref{ve}).

Next, it is a simple matter to check that
\begin{equation}
g_{t't'}=c^2 e^{-\lambda^2}
\label{gtM}
\end{equation}
which up to second order in $\lambda$ coincides with the ``Galilean''
result (\ref{gG}).
Also, we have in this case
\begin{equation}
g_{\phi'\phi'}=-\rho^2 e^{\lambda^2}\qquad ; \qquad
g_{t'\phi'}=0
\label{gfM}
\end{equation}
obtaining for the length of the circle of radius $\rho$
\begin{equation}
l'_{\phi} = 2 \pi \rho e^{\lambda^2/2}
\label{c}
\end{equation}
which up to terms of order $\lambda^2$ yields the desired expression
(\ref{ll2}).

Therefore MTT transformation disposes of objections previously raised
in conection with TT trasformations.

In what follows we shall present applications of MTT transformations
to some well known problems:

\subsection{The Sagnac effect}
This effect \cite{Sa},\cite{Po} consists in the phase shift
which appear between two counter propagating light beams on
a rotating platform.

Let us calculate this effect using MTT transformation. In the
rotating system $S'$ , the two rays which
left the point $A$ at the same time $t'_0$ return to $A$ at the
same time $t'_1=t'_2$ (however it is important to stress that the arrival
back at $A$ are two distinct point
events with different $\phi'$ coordinates in the rotating frame).

Now, the arrival of one of the rays back to $A$ is an event
with coordinates (in $S'$)
\begin{equation}
\rho'_1 \;\, , \,\; \phi'_1 = \phi'_0 + 2\pi \;\, , \,\; t'_1 = t'_0 +
\frac{2 \pi \rho_1 e^{\lambda^2/2}}{c}
\label{cop1}
\end{equation}
whereas the coordinates of the event consisting in the arrival
at $A$ of the other ray, are
\begin{equation}
\rho'_2=\rho'_1 \;\, , \,\; \phi'_2 = \phi'_0 - 2\pi \;\, , \,\; t'_2 = t'_1
\label{cop2}
\end{equation}
Therefore the difference of time in the arrival of both rays
at $A$, as measured by the observer in $S$, is given by
\begin{equation}
\Delta t = t_1 - t_2 = e^{\lambda^2/2} \left[\frac{4\pi}{c} \rho
\sinh{\lambda}\right]
\label{det}
\end{equation}
where (\ref{MTTt}) has been used.
Expanding (\ref{det}) up to order $\lambda^3$, we obtain
\begin{equation}
\Delta t = \frac{4\pi\omega\rho^2}{c^2} +
\frac{8\pi\omega^3\rho^4}{3c^4} + {\cal O}(\lambda^4)
\label{detex}
\end{equation}
The first term on the right of (\ref{detex}) is the
Sagnac's result. The second term is a correction of
order $\lambda^3$ introduced by our transformation. In the case of
TT transformation this correction is
\begin{equation}
\frac{2\pi\omega^3\rho^4}{3c^4}
\label{TTcorr}
\end{equation}

\subsection{Proper time intervals in $S$ and $S'$}
Let us now consider two events which take place at the same spatial
point as seen by $S'$. Thus, let $(z'_1,\rho'_1,\phi'_1,t'_1)$ and
$(z'_1,\rho'_1,\phi'_1,t'_2)$ be the coordinates of both events.
Then using (\ref{MTTt}), we have
\begin{equation}
\Delta t = t_2-t_1 = e^{-\lambda^2/2} \cosh{\lambda} \left(t'_2-t'_1\right)
\label{vt}
\end{equation}
Since $\Delta t'$ and $\Delta \tau'$ are related by (\ref{tautp}),
and $\Delta \tau=\Delta t$, we have
\begin{equation}
\Delta \tau=\Delta \tau' \cosh{\lambda}
\label{dta}
\end{equation}or, up to order $\lambda^2$
\begin{equation}
\Delta \tau = \Delta \tau' \left(1+\frac{\omega^2 \rho^2}{c^2}\right)
\label{dtau}
\end{equation}

\subsection{Simultaneous events in $S'$ and portable clocks}
Let us consider two synchronized clocks ($A$ and $B$) in $S'$, on
different points of the same parallel.

Then, two events with the same coordinate time $t'$ (simultaneous in $S'$)
at $A$ and $B$ have for the inertial observer $S$,
time coordinates which differ by
\begin{equation}
t_B-t_A = e^{\lambda^2/2} \frac{\rho}{c} \sinh{\lambda}
\left(\phi'_B-\phi'_A\right)
\label{dift}
\end{equation}
Let us now assume that $A$ and $B$ are at the same point in $S'$
showing the same time for both observers ($S$ and $S'$), then let us
rotate $B$ clockwise by $2\pi$ around the axis of rotation in $S'$.
When $B$ coincides again with $A$, the rotating observer will still read
the same time by both clocks, but for the inertial observer, $B$
will be slow with respect to $A$ by
\begin{equation}
\Delta t = e^{\lambda^2/2} 2\pi \frac{\rho}{c} \sinh{\frac{\omega\rho}{c}}
\label{sl}
\end{equation}
This result differs from the obtained with the TT transformation by
the exponential factor.

\section{Conclusions}
The TT transformation, which restores the most relevant aspects
of relativistic kinematics, has been reviewed.

To eliminate some undesirable consequences derived from its
application we have proposed a new set of transformations
which, while conserving all the advantages of TT, is free
from its most conspicuous drawbacks.

Some simple applications as the Sagnac's effect exhibit the differences
with TT and the ``Galilean'' transformation. These differences
being of order $\lambda^3$ and higher, it is not clear if they could
be detected with available technology.

However, for future navigation and time tranfer systems which
are now being contemplated, with sub-nanosecond time accuracy,
such terms could become relevant.

Although we do not claim that our proposition solves the
problem considered here in a definitive way, we do hope that
it will stimulate further discussions on this very important issue.

\end{document}